\RequirePackage{fix-cm}
\documentclass[smallextended]{svjour3}       
\smartqed  
\usepackage{graphicx}
\usepackage{undertilde}
\usepackage{color}
\begin{document}

\title{Conservation laws of an electro-active polymer
}


\author{Mireille Tixier         \and
        Jo\"{e}l Pouget 
}


\institute{M. Tixier \at
              Laboratoire d'Ing\'{e}nierie des Syst\`{e}mes de Versailles, Universit\'{e} de Versailles Saint Quentin, 45, avenue des Etats-Unis, F-78035 Versailles \\
              Tel.: +33-139254519\\
              Fax: +33-139254523\\
              \email{mireille.tixier@uvsq.fr}           
           \and
           J. Pouget \at
              Universit\'{e} Pierre et Marie Curie, UMR 7190, Institut Jean le Rond d'Alembert, F-75005 Paris, France \\
              CNRS, UMR 7190,  Institut Jean le Rond d'Alembert, F-75005 Paris, France \\
              Tel.: +33-144275465\\
              \email{pouget@lmm.jussieu.fr}           
}

\date{Received: date / Accepted: date}

\maketitle

\begin{abstract}
Ionic electro-active polymers (E.A.P.) is an active material consisting in 
a polyelectrolyte (for example Nafion). Such material is usually used as thin 
film sandwiched between two platinum electrodes. The 
polymer undergoes large bending motions when an electric field is applied
across the thickness. Conversely, a voltage can be detected between both
electrodes when the polymer is suddenly bent. The solvent-saturated polymer
is fully dissociated, releasing cations of small size. We used a continuous
medium approach. The material is modelled by the coexistence of two phases;
it can be considered as a porous medium where the deformable solid phase is
the polymer backbone with fixed anions; the electrolyte phase is made of a
solvent (usually water) with free cations.

The microscale conservation laws of mass, linear momentum and energy and the
Maxwell's equations are first written for each phase. The physical
quantities linked to the interfaces are deduced. The use of an average
technique applied to the two-phase medium finally leads to an Eulerian
formulation of the conservation laws of the complete material. Macroscale
equations relative to each phase provides exchanges through the interfaces.
An analysis of the balance equations of kinetic, potential and internal
energy highlights the phenomena responsible of the conversion of one kind of
energy into another, especially the dissipative ones : viscous frictions and
Joule effect.

\keywords{Electro-active polymers \and balance laws \and conservation laws \and multiphysics
coupling \and deformable porous media}
\PACS{PACS 47.10.ab \and PACS 47.56.+r \and PACS 61.41.+e \and PACS 83.60.Np}
\end{abstract}


\section{Introduction}

\label{intro} Electro-active polymers (EAP) have attracted much attention
from scientists and engineers of various disciplines. In particular,
researches in the field of biomimetics (for instance, in robotic mechanisms
are based on biologically-inspired models) and for the use as artificial
muscles (see, for instance, the review of Shahinpoor \cite{shahinpoor1998-a}
and \cite{shahinpoor1998-b} or \cite{samatham2007}) and more recently EAPs
are excellent candidates for energy harvesting devices \cite{sodano2004}, 
\cite{liu2004} and \cite{liu2005}. Roughly speaking, such polymers have
responses to external electric stimulation by displaying a significant
shape and size variations. This interesting property offers many promising
applications in advanced technologies. In addition, they can be used as
actuators or sensors. As actuators the EAPs are characterized by the fact
they undergo a large amount of deformation while sustaining large forces.
They are often called artificial muscles \cite{shahinpoor1994}, \cite%
{shahinpoor2000} and \cite{bar-cohen2002}. \newline
Electro-active polymers can be divided in several categories according to
their process of activation and chemical compositions. Nevertheless, they
can be placed in two major categories : electronic and ionic categories.
These both categories come in several families \cite{samatham2007} (among
them, ferroelectric polymers, dielectric EAP, electrostrictive paper,
electro-viscoelastic elastomers, ionic polymer gels, conductive polymers,
etc.). The first category of EAP is the electronic type. Concerning their
advantages the E.A.P. can operate in room conditions with rapid response in
time; in addition they induce relatively large actuation forces. One of the
main disadvantage is that they require high voltage ($150$ MV/m). The second
category, the ionic EAPs, with which the present work is concerned,
operates with low voltage (few volts) producing large bending displacements.
Their drawbacks are more or less slow response and low actuation force. They
operate best in humid environment and they can be made as self-contained
encapsulated actuators to be used in dry environment. \newline

In the present study the emphasis is placed especially on the ionic polymer
metal composite (IPMC) \cite{nemat2000}. The structure consists of thin
ion-exchange membrane of Nafion, Flemion or Aciplex (polyelectrolyte) plated
on both faces by conductive electrodes (generally platinum or gold). In
short, to explain the mechanism of deformation of an EAP, a thin trip of
polymers is placed between thin conductive electrodes. Upon the application
of an electric field across a slightly humid EAP, the positive counter ions
move towards the negative electrode (cathode), whole negative ions that are
fixed (or immobile) to the polymer backbone experience an attractive force
from the positive electrode (anode). At the same time, water molecules in
the EAP backbone diffuse towards the region of high positive ion
concentration (near the negative electrode) to equalize the charge
distribution. As a result, the water or solvent concentration in the region
near the anode increases and the concentration in the region near the
cathode decreases, leading to strain with linear distribution along the step
thickness which causes the bending towards the positive anode. Conversely,
if the strip of electro-active polymers is suddenly bent, a difference of
electric voltage is produced between electrodes \cite{newbury2002} and \cite%
{yoon2007}.\newline

The theories or models to explain the mechanism of deformation in EAP are yet
to emerge. Nevertheless, some heuristic or empiric models are available in
the literature. One of the most interesting and comprehensive accounts for
chemical mechano-electric effect of the ionic transport coupled to electric
field and elastic deformation of the polymer. A micro mechanical model has
been developped by Nemat-Nasser \cite{nemat2000} and \cite{nemat2002}
accounting for coupled ion transport, electric field and elastic deformation
to predict the response of the IPMC. The model presented is mostly governed
by Gauss equation for the conservation of electric charge, a constitutive
equation for ion flux vector and a so-called generalized Darcy's law for the
water molecule velocity. Other models based on linear irreversible
thermodynamics have been proposed by Shahinpoor \textit{et al.} \cite%
{shahinpoor2000} and \cite{degennes2000}. The model considers standard
Onsager formulation for the simple description of ion transport (current
density) and the flux of the solvent transport. The conjugate forces are the
electric field and the gradient of pressure. In different way, Shahinpoor
and co-workers propose models for micro-electro-mechanics of ions polymeric
gels based on continuum electromechanics \cite{shahinpoor1994}. \newline

The present work focus on a novel approach for electro-active polymers based
on thermodynamics of continua. More precisely, we present a detailed
approach for such polymer material using the concepts of non-equilibrium
thermodynamical processes. The material is then modeled by the coexistence
of two phases. The first one is the backbone polymer or the solid phase with
fixed anion while the second phase is the solvent containing the free
cations. The method consists of computing an average of the different phases
over a representative elementary volume containing the phases at the micro
scale. The statistical average leads to macro scale quantities defined all
over the material. The main difficulty of the method is that we must account
for the interfaces which exist between phases for which interfacial
quantities must be defined. On using this procedure for different
conservation laws of the present multiphase material, we deduce the equation
of mass conservation, the electric charge conservation, the conservation of
the momentum, the different energy balance equations at the macroscopic
scale of the whole material. \newline

The paper is organized as follows. The description of the model and the
definition of phases are presented with underlying physics in the next
Section. Section 3 is devoted to the equations of conservation of mass.
Since the polymer contains electric charges, the electric charge
conservation and interface equations are presented in Section 4. The
following Section places the emphasis on the linear momentum balance
equation where the macroscopic stress tensor is defined. Moreover, the
Maxwell's tensor is placed in evidence due to the action of the electric
field on the moving electric charges. The Section 6 presents the energy
balance laws, that is, the potential energy, the kinetic energy, the total
energy and internal energy balance equations. At last, the discussion is
reported in Section 7 and finally conclusions are drawn in Section 8.


\section{Modelling}

\label{sec:1} As mentioned in the introduction, the system under study is
made of a thin membrane of an ionic electro-active polymer saturated with
water and coated on both sides with thin metal layers used as electrodes.
Water, even in small quantity, causes a quasi-complete dissociation of the polymer and the release of
positive ions (cations) in water; negative ions (anions) remain bound to the
polymer backbone \cite{Chabe}. When an electric field perpendicular to the electrodes is
applied, cations move towards the negative side, carrying solvent away by an
osmosis phenomenon. This solvent displacement leads to a polymer swelling on
the negative electrode side and to a compression on the opposite side,
resulting in a bending of the strip.

To model this system, we describe the polymer chains as a deformable porous
medium; this solid is saturated by an ionic solution composed by water and
cations. The whole material is considered as a continuum, which is the
superposition of three systems whose velocity fields are different : a
deformable solid component made up of polymer backbone negatively charged and
fluid trapped in the unconnected porosity (the "solid component"), and a liquid
composed of water and cations located in the connected porosity. Anions are
bound to the solid component. Quantities relative to the different components will
be respectively identified by subscripts $1$, $2$ and $3$ for cations, solvent
and solid. Subscript $4$ will refer to the solution, i.e. both components $1$
and $2$. Quantities without subscript refer to the whole material. Solid and
solution are separated by an interface (subscript $i$) whose thickness is
supposed to be negligible. Components $2$, $3$ and $4$ as well as the global 
material are assimilated to continua. Modelling of the interface is
detailled in the appendix.

Solid and solution are supposed to be incompressible phases. We assume the
gravity and the magnetic field are negligible; the only external force
acting on the system is the electric force.

To describe this complex dispersed medium, we use a coarse-grained model
developed by Nigmatulin \cite{Nigmatulin79}, \cite{Nigmatulin90}, Drew \cite%
{Drew83}, Drew and Passman \cite{Drew98} and Ishii and Hibiki \cite{Ishii06}
for two-phase mixtures \cite{Lhuillier03}. We use two scales. The microscopic scale must be small enough so that the corresponding volume only contains a single phase (3 or 4), but large enough to use a continuous medium model. For Nafion completely saturated with water, it is about hundred Angstroms. At the macroscopic scale, the
representative elementary volume (R.E.V.) contains phases 3 and 4. It must be large enough so that average quantities relative to the whole material make sense, and small enough so that these quantities can be considered as local. Its characteristic length is about micron \cite{Colette}, \cite{Gierke} and \cite{Chabe}. For each phase $3$ and $4$, we
define a microscale Heaviside-like function of presence $\chi _{k}\left( 
\overrightarrow{r},t\right) $ by

\begin{equation}
\chi _{k}=1\hbox{ when phase }k\hbox{ occupies point }\overrightarrow{r}%
\hbox{ at time t,}\quad \chi _{k}=0\hbox{ otherwise}  \label{Presence}
\end{equation}%
$\chi _{k}$ remains unchanged in case of displacement following the
interface velocity $\overrightarrow{V_{i}^{0}}$. We obtain

\begin{equation}
\overrightarrow{grad}\chi _{k}=-\overrightarrow{n_{k}}\chi _{i}\qquad \frac{%
\partial \chi _{k}}{\partial t}=\overrightarrow{V_{i}^{0}}\cdot\overrightarrow{%
n_{k}}\chi _{i}\qquad \mbox{for}\;k=3,4  \label{qui}
\end{equation}%
where the Dirac-like function $\chi _{i}=-\overrightarrow{grad}\chi _{k}\cdot%
\overrightarrow{n_{k}}$ (in $m^{-1}$) denotes the function of presence of
the interface and $\overrightarrow{n_{k}}$ the outward-pointing unit normal
to the interface in the phase $k$.

The quantities related to each phase have significant variation over space
and time, as well as the positions of each phase. In order to define
macroscale quantities relative to the whole material, we consider a
representative elementary volume (R.E.V.) containing the three components and
the microscale quantities are statistically averaged over the R.E.V.. This
statistical average, denoted by $\left\langle {}\right\rangle $ and obtained
by repeating many times the same experiment with the same boundary and
initial conditions, is supposed to be equivalent to a volume average
(ergodic hypothesis). The average thus defined commutes with the space and
time derivatives (Leibniz' and Gauss' rules, Drew \cite{Drew83}; Lhuillier \cite{Lhuillier03}). On denoting by $\left\langle{}\right\rangle _{k}$ the average over the phase $k$ of a quantity relative
to the phase $k$ only, a microscale quantity $g_{k}^{0}$ satisfies

\begin{equation}
g_{k}=\left\langle \chi _{k}g_{k}^{0}\right\rangle =\phi _{k}\left\langle
g_{k}^{0}\right\rangle _{k}
\end{equation}%
where $\phi _{k}=\left\langle \chi _{k}\right\rangle $ is the volume
fraction of the phase $k$. The macroscale quantity $g_{k}$ is defined all
over the material. In the following, superscript $^{0}$ denotes the
microscale quantities of each phase. The macroscale quantities, which are
averages defined everywhere, are written without superscript.


\section{Equation of conservation of mass}

\label{sec:2} In the following, we assume that the polymer is enough hydrated to be completely dissociated. For the water, solution and solid phases, the microscale mass continuity equation can be written as

\begin{equation}
\frac{\partial \rho _{k}^{0}}{\partial t}+div\left( \rho _{k}^{0}%
\overrightarrow{V_{k}^{0}}\right) =0  \label{CMm}
\end{equation}%
where $\overrightarrow{V_{k}^{0}}$ is the local velocity of the phase $k$
and $\rho _{k}^{0}$ its mass density. Phases $2$ and $3$ are incompressible,
so we obtain

\begin{equation}
div\left( \overrightarrow{V_{k}^{0}}\right) =0  \label{CMmbis}
\end{equation}

The different phases do not interpenetrate, thus we can write

\begin{equation}
\overrightarrow{V_{1}^{0}}\chi _{i}=\overrightarrow{V_{2}^{0}}\chi _{i}=%
\overrightarrow{V_{3}^{0}}\chi _{i}=\overrightarrow{V_{4}^{0}}\chi _{i}=%
\overrightarrow{V_{i}^{0}}\chi _{i}  \label{CMcl}
\end{equation}

Using (\ref{CMm}) and (\ref{qui}) we deduce

\begin{equation}
\frac{\partial \chi _{k}\rho _{k}^{0}}{\partial t}+div\left( \chi _{k}\rho
_{k}^{0}\overrightarrow{V_{k}^{0}}\right) =\rho _{k}^{0}\overrightarrow{%
V_{i}^{0}}\cdot\overrightarrow{n_{k}}\chi _{i}-\rho _{k}^{0}\overrightarrow{%
V_{k}^{0}}\cdot\overrightarrow{n_{k}}\chi _{i}  \label{1}
\end{equation}

For the phase $k$, the mass density relative to the whole material volume
and the barycentric velocity are defined respectively by

\begin{equation}
\rho _{k}=\left\langle \chi _{k}\rho _{k}^{0}\right\rangle =\phi _{k}\rho
_{k}^{0}\qquad \qquad \overrightarrow{V_{k}}=\frac{\left\langle \chi
_{k}\rho _{k}^{0}\overrightarrow{V_{k}^{0}}\right\rangle }{\left\langle \chi
_{k}\rho _{k}^{0}\right\rangle }=\overrightarrow{V_{k}^{0}}
\end{equation}%
neglecting the velocities fluctuations on the R.E.V. scale.

\begin{equation}
\rho _{4}^{0}=\rho _{2}^{0}\frac{\phi _{2}}{\phi _{4}}+CM_{1}
\end{equation}%
where $M_{k}$ is the molar mass of the component $k$ and $C$ the cations molar
concentration relative to the solution volume. It follows%
\begin{equation}
\rho _{4}=\rho _{1}+\rho _{2}\qquad \mbox{with}\qquad \rho _{1}=\phi
_{4}CM_{1}
\end{equation}%
assuming that the concentration fluctuations are negligible and that the
solution is diluted. In the same way the velocity of the solution can be
written as

\begin{equation}
\rho _{4}^{0}\overrightarrow{V_{4}^{0}}=CM_{1}\overrightarrow{V_{1}^{0}}%
+\rho _{2}^{0}\frac{\phi _{2}}{\phi _{4}}\overrightarrow{V_{2}^{0}}\qquad
\qquad \rho _{4}\overrightarrow{V_{4}}=\rho _{1}\overrightarrow{V_{1}}+\rho
_{2}\overrightarrow{V_{2}}
\end{equation}

Averaging over the material R.E.V., we finally obtain%
\begin{equation}
\frac{\partial \rho _{k}}{\partial t}+div\left( \rho _{k}\overrightarrow{%
V_{k}}\right) =0\qquad \qquad k=1,2,3,4  \label{CMk}
\end{equation}

The interfaces have no mass. Consequently, we deduce for the complete
material

\begin{equation}
\frac{\partial \rho }{\partial t}+div\left( \rho \overrightarrow{V}\right) =0
\label{CM}
\end{equation}%
where $\rho $ and $\overrightarrow{V}$ denote the mass density and the
barycentric velocity of the whole material

\begin{equation}
\rho =\sum\limits_{k=3,4}\rho _{k}\qquad \qquad \rho \overrightarrow{V}%
=\sum\limits_{k=3,4}\rho _{k}\overrightarrow{V_{k}}
\end{equation}


\section{Electric equations}

\label{sec:3}

\subsection{Electric charge conservation}

\label{sec:31} The microscale electric charge conservation of the
phase $k$ can be written%
\begin{equation}
div\overrightarrow{I_{k}^{0}}+\frac{\partial \left( \rho
_{k}^{0}Z_{k}^{0}\right) }{\partial t}=0  \label{CCm}
\end{equation}%
where $\overrightarrow{I_{k}^{0}}$ denotes the current density vector and $%
Z_{k}^{0}$ the electric charge per unit of mass ($Z_{2}^{0}$ and $Z_{3}^{0}$
are constants).

\begin{equation}
\overrightarrow{I_{3}^{0}}=\rho _{3}^{0}Z_{3}^{0}\overrightarrow{V_{3}^{0}}%
\qquad \qquad \qquad \overrightarrow{I_{4}^{0}}=M_{1}CZ_{1}^{0}%
\overrightarrow{V_{1}^{0}}
\end{equation}

\begin{equation}
Z_{k}^{0}=\frac{z_{k}F}{M_{k}}\quad \quad \mbox{for}\quad k=1,3\qquad
\quad Z_{2}^{0}=0 \qquad \quad Z_{4}^{0}=\frac{CM_{1}Z_{1}^{0}}{\rho _{4}^{0}}
\end{equation}%
where $z_{k}$ is the number of elementary charges of an ion and $F$ the
Faraday's constant.

Averaging over the R.E.V., we obtain%
\begin{equation}
div\overrightarrow{I_{k}}+\frac{\partial \rho _{k}Z_{k}}{\partial t}%
=\left\langle -\overrightarrow{i_{k}^{0}}\cdot\overrightarrow{n_{k}}\chi
_{i}\right\rangle  \label{CCk}
\end{equation}%
in which the macroscale mass charge and current density vector are defined as

\begin{equation}
\rho _{k}Z_{k}=\left\langle \chi _{k}\rho _{k}^{0}Z_{k}^{0}\right\rangle
\qquad \qquad \overrightarrow{I_{k}}=\left\langle \chi _{k}\overrightarrow{%
I_{k}^{0}}\right\rangle
\end{equation}%
with%
\begin{equation}
\overrightarrow{I_{3}}=\left\langle \chi _{3}\overrightarrow{I_{3}^{0}}%
\right\rangle =\rho _{3}Z_{3}\overrightarrow{V_{3}}\qquad \qquad 
\overrightarrow{I_{4}}=\left\langle \chi _{4}\overrightarrow{I_{4}^{0}}%
\right\rangle =\rho _{1}Z_{1}\overrightarrow{V_{1}}
\end{equation}%
$\overrightarrow{i_{k}^{0}}=\overrightarrow{I_{k}^{0}}-\rho _{k}^{0}Z_{k}^{0}%
\overrightarrow{V_{k}^{0}}$ denotes the microscale diffusion current in
phase $k$. Quantities relative to the interfaces are defined in the appendix.
The interface electric charge density per unit surface $Z_{i}$ and the
current density vector $\overrightarrow{I_{i}}$ satisfy the following mean
condition%
\begin{equation}
\frac{\partial Z_{i}}{\partial t}+div\overrightarrow{I_{i}}=\left\langle 
\overrightarrow{i_{3}^{0}}\cdot\overrightarrow{n_{3}}\chi _{i}+\overrightarrow{%
i_{4}^{0}}\cdot\overrightarrow{n_{4}}\chi _{i}\right\rangle  \label{CCi}
\end{equation}

Adding up equations (\ref{CCk}) for the solid, the solution and (\ref{CCi}) for the
interfaces, it follows for the whole material%
\begin{equation}
div\overrightarrow{I}+\frac{\partial \rho Z}{\partial t}=0  \label{CC}
\end{equation}%
where%
\begin{equation}
\rho Z=\sum\limits_{3,4}\rho _{k}Z_{k}+Z_{i}\qquad \qquad \overrightarrow{I}%
=\rho _{1}Z_{1}\overrightarrow{V_{1}}+\rho _{3}Z_{3}\overrightarrow{V_{3}}+%
\overrightarrow{I_{i}}
\end{equation}

\subsection{Maxwell's equations}

\label{sec:32} One can reasonably neglect the effects of the magnetic field.
The electric fields $\overrightarrow{E_{k}^{0}}$ and the electric
displacements $\overrightarrow{D_{k}^{0}}$ of the solid and the solution are
governed by the Maxwell's equations

\begin{equation}
\overrightarrow{rot}\overrightarrow{E_{k}^{0}}=\overrightarrow{0}\qquad
\qquad div\overrightarrow{D_{k}^{0}}=\rho _{k}^{0}Z_{k}^{0}  \label{MAXm}
\end{equation}

The associated boundary conditions can be presented as%
\begin{equation}
\overrightarrow{n_{3}}\wedge \overrightarrow{E_{3}^{0}}\chi _{i}=-%
\overrightarrow{n_{4}}\wedge \overrightarrow{E_{4}^{0}}\chi _{i}\qquad
\qquad \ \ \overrightarrow{D_{3}^{0}}\cdot\overrightarrow{n_{3}}\chi _{i}+%
\overrightarrow{D_{4}^{0}}\cdot\overrightarrow{n_{4}}\chi _{i}+Z_{i}^{0}\chi
_{i}=0  \label{MAXcl}
\end{equation}

Averaging equations (\ref{MAXm}) over the R.E.V., we derive the following
macroscale equations for the solid and the solution%

\begin{equation}
\overrightarrow{rot}\overrightarrow{E_{k}}=\overrightarrow{0}\qquad \qquad
div\overrightarrow{D_{k}}=\rho _{k}Z_{k}-\left\langle \overrightarrow{%
D_{k}^{0}}\cdot\overrightarrow{n_{k}}\chi _{i}\right\rangle  \label{MAXk}
\end{equation}%
in which the macroscale electric fields and displacements are defined as%

\begin{equation}
\overrightarrow{E_{k}}=\frac{\left\langle \chi_{k} \overrightarrow{E_{k}^{0}}%
\right\rangle }{\left\langle \chi_{k} \right\rangle }\qquad \qquad 
\overrightarrow{D_{k}}=\left\langle \chi _{k} \overrightarrow{D_{k}^{0}}\right\rangle
\end{equation}%

Electric field is an intensive thermodynamic variable. In principle, it
displays spatial and time fluctuations within the R.E.V.. Considering this
volume is tiny, we assume that the fluctuations are not relevant; we venture
the same hypothesis for the concentration and the velocities of the phases.
Furthermore, we suppose that macroscale electric fields are identical in all
the phases. Adding up equations (\ref{MAXk}) for the solid and the solution,
it follows for the whole material%

\begin{equation}
\overrightarrow{rot}\overrightarrow{E}=\overrightarrow{0}\qquad \qquad div%
\overrightarrow{D}=\rho Z  \label{MAX}
\end{equation}%
using (\ref{MAXcl}). Parameters of the complete material are defined by%

\begin{equation}
\overrightarrow{E}=\sum\limits_{3,4}\phi _{k}\overrightarrow{E_{k}}=\overrightarrow{E_{k}}
\qquad\qquad \overrightarrow{D}=\sum\limits_{3,4}\overrightarrow{D_{k}}
\end{equation}

We conclude that the E.A.P. verifies the same Maxwell's equations and the
same law of conservation of charge as an isotropic homogeneous linear
dielectric.

\subsection{Constitutive relations}

\label{sec:33} A reasonable approximation is that solid and solution can be
regarded as isotropic linear dielectrics%

\begin{equation}
\overrightarrow{D_{k}^{0}}=\varepsilon _{k}^{0}\overrightarrow{E_{k}^{0}}
\label{RCm}
\end{equation}%
where $\varepsilon _{k}^{0}$ denotes the permittivity of the phase $k$.
Average over the R.E.V. gives%

\begin{equation}
\overrightarrow{D_{k}}=\varepsilon _{k}\overrightarrow{E_{k}}  \label{RCk}
\end{equation}%
in which :

\begin{equation}
\varepsilon _{k}=\left\langle \chi _{k}\varepsilon _{k}^{0}\right\rangle
\end{equation}%
is the mean permeability of the phase $k$ relative to the total volume.

The constitutive relation of the E.A.P. takes on the following form%

\begin{equation}
\overrightarrow{D}=\varepsilon \overrightarrow{E}  \label{RC}
\end{equation}%
where the whole material permittivity is defined by%

\begin{equation}
\varepsilon =\sum\limits_{k=3,4}\varepsilon _{k}
\end{equation}

On considering our assumptions, the E.A.P. is equivalent to an isotropic
linear dielectric. We however point out that its permittivity a priori
varies over time and space because of variations of the volume fractions $%
\phi _{3}$ and $\phi _{4}$.


\section{Linear momentum conservation law}

\label{sec:4}

\subsection{Particle derivatives and material derivative}

\label{sec:41} In order to write the remaining balance equations, it is
necessary to calculate the variations of the extensive quantities following
the material motion. This raises a problem because the different phases
do not move with the same velocity : velocities of the solid and the solution
are a priori different. For a quantity $g$, we can define
particle derivatives following the motion of the solid $(\frac{d_{3}}{dt})$%
, the solution $(\frac{d_{4}}{dt})$ or the interface $(\frac{d_{i}}{dt})$

\begin{equation}
\frac{d_{k}g}{dt}=\frac{\partial g}{\partial t}+\overrightarrow{grad}g\cdot%
\overrightarrow{V_{k}}
\end{equation}

Let us consider an extensive quantity of density $g\left( \overrightarrow{r}%
,t\right) $ relative to the whole material. According to the theory
developped by O. Coussy \cite{Coussy95} and implicitly used in \cite{Biot77}
and \cite{Coussy89}, we are able to define a derivative following the motion
of the different phases of the medium. We will call it the "material
derivative"

\begin{equation}
\frac{D}{Dt}\left( \frac{g}{\rho }\right) =\sum\limits_{k=3,4,i}\frac{\rho
_{k}}{\rho }\frac{d_{k}\left( \frac{g_{k}}{\rho _{k}}\right) }{dt}
\end{equation}%
where $g_{3}$, $g_{4}$ and $g_{i}$ are the densities relative to the total
actual volume attached to the solid, the solution and the interface,
respectively (for example, if $g$ is the volume density, we set $%
g_{3}=1-\phi $ and $g_{4}=\phi $ where $\phi $ is the porosity)

\begin{equation}
g=g_{3}+g_{4}+g_{i}
\end{equation}%
$\frac{d_{k}}{dt}\left( \frac{g_{k}}{\rho _{k}}\right) $ is the derivative
following the motion of the phase $k$ of the mass density associated with
the quantity $g_{k}$. Using (\ref{CMk}), we derive

\begin{equation}
\rho \frac{D\left( \frac{g}{\rho }\right) }{Dt}=\sum\limits_{k=3,4,i}\frac{%
\partial g_{k}}{\partial t}+div\left( g_{k}\overrightarrow{V_{k}}\right)
\label{DerivMat}
\end{equation}%
for a scalar quantity and

\begin{equation}
\rho \frac{D\left( \frac{\overrightarrow{g}}{\rho }\right) }{Dt}%
=\sum\limits_{k=3,4,i}\frac{\partial \overrightarrow{g_{k}}}{\partial t}%
+\overrightarrow{div}\left( \overrightarrow{g_{k}}\otimes \overrightarrow{V_{k}}\right)
\label{DerivMatVect}
\end{equation}%
for a vector quantity. This derivative must not be confused with the
derivative $\frac{d}{dt}$ following the barycentric velocity $%
\overrightarrow{V}$.

\subsection{Linear momentum balance equation}

\label{sec:42} On assuming that the gravity and the magnetic field are
negligible, the only applied volume force is the electric one. The
microscale momentum balance equation of the phase $k$ is then written as

\begin{equation}
\frac{\partial \rho _{k}^{0}\overrightarrow{V_{k}^{0}}}{\partial t}%
+\overrightarrow{div}\left( \rho _{k}^{0}\overrightarrow{V_{k}^{0}}\otimes \overrightarrow{%
V_{k}^{0}}\right) =\overrightarrow{div}\utilde{\sigma} _{k}^{0}+\rho
_{k}^{0}Z_{k}^{0}\overrightarrow{E_{k}^{0}}  \label{CQm}
\end{equation}%

where $\utilde{\sigma} _{k}^{0}$, the microscale
stress tensor of the phase $k$, is symmetric. The linear momentum of the
interfaces per surface unit is zero (see appendix). On accounting for the
assumptions concerning the local velocities, it follows that at the macroscopic
scale

\begin{equation}
\frac{\partial \rho _{k}\overrightarrow{V_{k}}}{\partial t}+\overrightarrow{%
div}\left( \rho _{k}\overrightarrow{V_{k}}\otimes \overrightarrow{V_{k}}%
\right) =\overrightarrow{div}\utilde{\sigma} _{k}%
+\rho _{k}Z_{k}\overrightarrow{E_{k}}+\overrightarrow{F_{k}}  \label{CQk}
\end{equation}%
where

\begin{equation}
\utilde{\sigma}_{k}=\left\langle \chi _{k}%
\utilde{\sigma}_{k}^{0} \right\rangle \qquad \qquad 
\overrightarrow{F_{k}}=\left\langle \utilde{\sigma}
_{k}^{0} \cdot\overrightarrow{n_{k}}\chi _{i}\right\rangle
\end{equation}

We verify that the macroscale stress tensor of the phase $k$, $%
\utilde{\sigma} _{k}$, is symmetric. $%
\overrightarrow{F_{k}}$ represents the resultant of the mechanical stresses
exerted on the phase $k$ by the other phase; it is an interaction force.
Concerning the interfaces, we obtain the following mean condition (cf \S\ %
\ref{sec:Annexe}), which expresses\ the linear momentum conservation law for
the interfaces

\begin{equation}
\overrightarrow{F_{3}}+\overrightarrow{F_{4}}=Z_{i}\overrightarrow{E_{i}}
\label{CQcl}
\end{equation}

The interface momentum is zero, then the volume linear momentum of the
whole material is $\rho \overrightarrow{V}$ $=\rho _{3}\overrightarrow{V_{3}}%
+\rho _{4}\overrightarrow{V_{4}}$. On using the definition of the material
derivative (\ref{DerivMatVect}), we obtain

\begin{equation}
\rho \frac{D\overrightarrow{V}}{Dt}=\overrightarrow{div}\utilde{\sigma }%
+\rho Z\overrightarrow{E}  \label{CQ}
\end{equation}%
in which

\begin{equation}
\utilde{\sigma}=\sum\limits_{k=3,4}\utilde{\sigma}_{k}
\end{equation}

We check that $\utilde{\sigma }$ is a symmetric
tensor and that in the absence of any external force ($\overrightarrow{E}=%
\overrightarrow{0}$), the total linear momentum is conserved.

Using Maxwell's equations (\ref{MAX}) and (\ref{RC}), (\ref{CQ}) becomes

\begin{equation}
\rho \frac{D\overrightarrow{V}}{Dt}=\overrightarrow{div}\left[ 
\utilde{\sigma }+\varepsilon \left( \overrightarrow{E}\otimes \overrightarrow{E}%
-\frac{E^{2}}{2}\utilde{I}\right) \right] +\frac{E^{2}}{2}\overrightarrow{grad}\varepsilon
\end{equation}%
$\varepsilon \left( \overrightarrow{E}\otimes \overrightarrow{E}-\frac{E^{2}%
}{2}\utilde{I}\right) $ is the Maxwell's tensor, which is here symmetric. The 
additional term $\frac{E^{2}}{2}\overrightarrow{grad}\varepsilon $ is produced 
by the non homogeneous material permittivity.

\section{Energy balance laws}

\label{sec:5}

\subsection{Potential energy balance equation}

\label{sec:51} Solid and solution are supposed to be non-dissipative
isotropic linear media. As a consequence the balance equation for the potential energy or
Poynting's theorem can be written in the integral form \cite{Jackson}, \cite{Maugin}%
\begin{equation}
\frac{d}{dt}\int_{\Omega }\frac{1}{2}\left( \overrightarrow{E}\cdot\overrightarrow{D}+%
\overrightarrow{B}\cdot\overrightarrow{H}\right) dv=-\oint\nolimits_{\partial
\Omega }\left( \overrightarrow{E}\wedge \overrightarrow{H}\right) \cdot%
\overrightarrow{n}ds-\int_{\Omega }\overrightarrow{E}\cdot\overrightarrow{I}dv
\end{equation}%
assuming that no charge goes out of the volume $\Omega $. The left hand side
represents the variation of the potential energy attached to the volume $%
\Omega $ following the charge motion. If the charges are mobile, the
associated local equation writes for the phase $k$, neglecting the magnetic
field

\begin{equation}
\frac{\partial E_{pk}^{0}}{\partial t}+div\left( E_{pk}^{0}\overrightarrow{%
V_{k}^{0}}\right) =-\overrightarrow{E_{k}^{0}}\cdot\overrightarrow{I_{k}^{0}}%
\qquad \qquad k=3,4  \label{Epm}
\end{equation}%
in which

\begin{equation}
E_{pk}^{0}=\frac{1}{2}\overrightarrow{D_{k}^{0}}\cdot\overrightarrow{E_{k}^{0}}%
\qquad \qquad k=3,4
\end{equation}%
is the potential energy per unit of volume of the phase $k$. On taking the
statistical average of (\ref{Epm}) over the R.E.V., we obtain

\begin{equation}
\frac{\partial E_{pk}}{\partial t}+div\left( E_{pk}\overrightarrow{V_{k}}%
\right) =-\overrightarrow{E_{k}}\cdot\overrightarrow{I_{k}}  \label{Epk}
\end{equation}%
where

\begin{equation}
E_{pk}=\left\langle \chi _{k}E_{pk}^{0}\right\rangle =\frac{1}{2}%
\overrightarrow{D_{k}}\cdot\overrightarrow{E_{k}}
\end{equation}

The mean volume potential energy associated to the interfaces satisfies
(see appendix)

\begin{equation}
\frac{\partial E_{pi}}{\partial t}+div\left( E_{pi}\overrightarrow{V_{i}}%
\right) =-\overrightarrow{E_{i}}\cdot\overrightarrow{I_{i}}
\end{equation}

The potential energy balance equation for the whole material is then

\begin{equation}
\rho \frac{D}{Dt}\left( \frac{E_{p}}{\rho }\right) =-\overrightarrow{E}\cdot%
\overrightarrow{I}  \label{Ep}
\end{equation}

where

\begin{equation}
E_{p}=\sum_{3,4,i}E_{pi}=\frac{1}{2}\overrightarrow{D}\cdot\overrightarrow{E}
\end{equation}

The production of potential energy in the R.E.V. is equal to the volume
power $-\overrightarrow{E}\cdot\overrightarrow{I}$ of the force due to the
action of the electric field on the density of electric charges.

\subsection{Kinetic energy balance equation}

\label{sec:52} The microscale kinetic energy balance equation derives from (%
\ref{CQm})

\begin{equation}
\frac{\partial E_{ck}^{0}}{\partial t}+div\left( E_{ck}^{0}\overrightarrow{%
V_{k}^{0}}\right) =div\left( \utilde{\sigma} _{k}^{0}\overrightarrow{V_{k}^{0}}%
\right) -\utilde{\sigma}_{k}^{0}:\utilde{grad}\overrightarrow{V_{k}^{0}}%
+\rho _{k}^{0}Z_{k}^{0}\overrightarrow{E_{k}^{0}}\cdot\overrightarrow{V_{k}^{0}}
\label{Ecm}
\end{equation}%
where the microscale volume kinetic energy of the phase $k$ is

\begin{equation}
E_{ck}^{0}=\frac{1}{2}\rho _{k}^{0}V_{k}^{02}
\end{equation}

In the same way, (\ref{CQk}) is transformed into

\begin{equation}
\frac{\partial E_{ck}}{\partial t}+div\left( E_{ck}\overrightarrow{V_{k}}%
\right) =\overrightarrow{V_{k}}\cdot\overrightarrow{div}\utilde{\sigma}_{k}+\rho _{k}Z_{k}%
\overrightarrow{V_{k}}\cdot\overrightarrow{E_{k}}+%
\overrightarrow{F_{k}}\cdot\overrightarrow{V_{k}}  \label{Eck}
\end{equation}%
where

\begin{equation}
E_{ck}=\frac{1}{2}\rho _{k}V_{k}^{2}
\end{equation}%
is the macroscale volume kinetic energy of the phase $k$. The interface
kinetic energy \ is zero (see appendix).

On summing up the equations (\ref{Eck}) for phases $3$ and $4$, we arrive at

\begin{equation}
\begin{tabular}{ll}
$\rho \frac{D}{Dt}\left( \frac{E_{c\Sigma }}{\rho }\right) $ & $%
=\sum\limits_{3,4}\left[ \frac{\partial E_{ck}}{\partial t}+div\left( E_{ck}%
\overrightarrow{V_{k}}\right) \right] $ \\ 
& $=\sum\limits_{3,4}\left[ div\left( \utilde{\sigma}
_{k}\cdot\overrightarrow{V_{k}}\right) -\utilde{\sigma}
_{k}:\utilde{grad}\overrightarrow{V_{k}}\right] +%
\left[ \sum\limits_{3,4}\rho _{k}Z_{k}\overrightarrow{V_{k}}+Z_{i}%
\overrightarrow{V_{i}}\right] \cdot\overrightarrow{E}$%
\end{tabular}
\label{Ecsigma}
\end{equation}%
where $E_{c\Sigma }$ is the sum of the volume kinetic energies of the
different phases with respect to the laboratory reference frame

\begin{equation}
E_{c\Sigma }=E_{c3}+E_{c4}
\end{equation}%
$\qquad E_{c\Sigma }$ is distinct from the kinetic energy of the whole
material because the phase velocities are different. The total volume
kinetic energy $E_{c}$ is defined as

\begin{equation}
E_{c}=\frac{1}{2}\rho V^{2}=\sum\limits_{3,4}\frac{1}{2}\rho _{k}V^{2}
\end{equation}%
From (\ref{DerivMat}), we deduce

\begin{equation}
\rho \frac{D}{Dt}\left( \frac{E_{c}}{\rho }\right) =\frac{\partial E_{c}}{%
\partial t}+div\left( E_{c}\overrightarrow{V}\right)
\end{equation}%
Using (\ref{Ecsigma}), it follows

\begin{equation}
\begin{tabular}{l}
$\rho \frac{D}{Dt}\left( \frac{E_{c}}{\rho }\right) =\frac{\partial }{%
\partial t}\left( E_{c}-\sum\limits_{3,4}E_{ck}\right) +div\left[
\sum\limits_{3,4}\left( \utilde{\sigma}_{k}\cdot%
\overrightarrow{V_{k}}-E_{ck}\overrightarrow{V_{k}}\right) +E_{c}%
\overrightarrow{V}\right] $ \\ 
$\qquad \qquad \qquad -\sum\limits_{3,4}\utilde{\sigma}_{k}:\utilde{grad}%
\overrightarrow{V_{k}}+\left( \sum\limits_{3,4}\rho _{k}Z_{k}\overrightarrow{V_{k}}%
+Z_{i}\overrightarrow{V_{i}}\right) \cdot\overrightarrow{E}$%
\end{tabular}
\label{Ec}
\end{equation}%
The last two terms of this equation are source terms. The penultimate one
represents the viscous dissipation, that is to say kinetic energy conversion
into internal energy. The last term is the electric force volume power,
which corresponds to a potential energy conversion into kinetic energy. As
for the first two terms, they correspond to the kinetic energy flux, which
is both due to the contact forces work $\sum\limits_{3,4}div\left( 
\utilde{\sigma}_{k}\cdot\overrightarrow{V_{k}}\right) $
and to the relative velocity of the two phases: the kinetic energy of the
phases with respect to the barycentric reference frame becomes indeed part
of the internal energy of the whole material.

\subsection{Total energy conservation law}

\label{sec:53} The total energy of the present system is the sum of its
internal, potential and kinetic energies. The energy fluxes come from
contact forces work and heat conduction. The microscale energy conservation
law for the phase $k$ can be written as

\begin{equation}
\frac{\partial E_{k}^{0}}{\partial t}+div\left[ E_{k}^{0}\overrightarrow{%
V_{k}^{0}}-\utilde{\sigma}_{k}^{0}\cdot\overrightarrow{%
V_{k}^{0}}+\overrightarrow{Q_{k}^{0}}\right] =0  \label{Em}
\end{equation}%
where%
\begin{equation}
E_{k}^{0}=U_{k}^{0}+\frac{1}{2}\rho _{k}^{0}V_{k}^{02}+\frac{1}{2}%
\overrightarrow{E_{k}^{0}}\cdot\overrightarrow{D_{k}^{0}} 
\end{equation}%
is the total microscale energy of the phase $k$. $\overrightarrow{Q_{k}^{0}}$
denotes the microscale heat flux of the phase $k$ and $U_{k}^{0}$ its
microscale internal energy. Average over the R.E.V. leads to

\begin{equation}
\frac{\partial E_{k}}{\partial t}+div\left( E_{k}\overrightarrow{V_{k}}%
\right) -div\left( \utilde{\sigma}_{k}\cdot%
\overrightarrow{V_{k}}\right) +div\overrightarrow{Q_{k}}=\overrightarrow{%
F_{k}}\cdot\overrightarrow{V_{k}}+P_{k}  \label{Ek}
\end{equation}%
where

\begin{equation}
E_{k}=\left\langle \chi _{k}E_{k}^{0}\right\rangle
=U_{k}+E_{ck}+E_{pk}\qquad \quad U_{k}=\left\langle \chi
_{k}U_{k}^{0}\right\rangle \qquad \quad \overrightarrow{Q_{k}}=\left\langle
\chi _{k}\overrightarrow{Q_{k}^{0}}\right\rangle
\end{equation}%
and

\begin{equation}
P_{k}=\left\langle -\overrightarrow{Q_{k}^{0}}\cdot\overrightarrow{n_{k}}\chi
_{i}\right\rangle
\end{equation}

$\overrightarrow{F_{k}}\cdot\overrightarrow{V_{k}}+P_{k}$ represents the energy
exchanges between the different phases through the interfaces : contact
forces work and heat fluxes. We obtain the following condition for the
interfaces (see appendix)

\begin{equation}
\frac{\partial E_{i}}{\partial t}+div\left( E_{i}\overrightarrow{V_{i}}%
\right) =-P_{3}-P_{4}-\overrightarrow{F_{3}}\cdot\overrightarrow{V_{3}}-%
\overrightarrow{F_{4}}\cdot\overrightarrow{V_{4}}  \label{Ei}
\end{equation}%
where $E_{i}$ is the total energy density of the interfaces averaged over
the R.E.V.. On summing equations (\ref{Ek}) for $k=3,4$ and (\ref{Ei}), we
obtain the conservation law of the total volume energy of the whole material 
$E$

\begin{equation}
\rho \frac{D}{Dt}\left( \frac{E}{\rho }\right) =div\left( \sum\limits_{k=3,4}%
\utilde{\sigma}_{k}\cdot\overrightarrow{V_{k}}\right)
-div\overrightarrow{Q}  \label{E}
\end{equation}%
where

\begin{equation}
E=\sum\limits_{3,4,i}E_{k}=U+E_{c}+E_{p}\quad \quad \quad \quad \quad \quad
\quad \overrightarrow{Q}=\sum\limits_{k=3,4}\overrightarrow{Q_{k}}
\end{equation}

The source term of this equation is zero, which is the expression of the
conservation law of the energy. $\sum\limits_{3,4}\utilde{\sigma}_{k}\cdot%
\overrightarrow{V_{k}}$ and $\overrightarrow{Q}$ represent the volume power of 
the contact forces and the heat fluxes of the complete medium, respectively.

\subsection{Internal energy balance equation}

\label{sec:54} The internal energy equation is obtained by subtracting
kinetic and potential energy equations (\ref{Ecm}) and (\ref{Epm})\ from the 
total energy conservation law (\ref{Em})

\begin{equation}
\frac{\partial U_{k}^{0}}{\partial t}+div\left( U_{k}^{0}\overrightarrow{%
V_{k}^{0}}+\overrightarrow{Q_{k}^{0}}\right) =\utilde{\sigma}_{k}^{0}:\utilde{grad}%
\overrightarrow{V_{k}^{0}}+\left( \overrightarrow{I_{k}^{0}}-\rho
_{k}^{0}Z_{k}^{0}\overrightarrow{V_{k}^{0}}\right) \cdot\overrightarrow{E_{k}^{0}%
}  \label{Um}
\end{equation}

Algebraic manipulations of (\ref{Ek}), (\ref{Eck}) and (\ref{Epk}) lead to

\begin{equation}
\frac{\partial U_{k}}{\partial t}+div\left( U_{k}\overrightarrow{V_{k}}+%
\overrightarrow{Q_{k}}\right) =\utilde{\sigma}_{k}:\utilde{grad}\overrightarrow{V_{k}}%
+ \overrightarrow{i_{k}}\cdot\overrightarrow{E_{k}}-\left\langle \overrightarrow{Q_{k}^{0}}%
\cdot\overrightarrow{n_{k}}\chi _{i}\right\rangle   \label{Uk}
\end{equation}%
and for the interfaces (see appendix)

\begin{equation}
\frac{\partial U_{i}}{\partial t}+div\left( U_{i}\overrightarrow{V_{i}}%
\right) =\left\langle \overrightarrow{Q_{3}^{0}}\cdot\overrightarrow{n_{3}}\chi
_{i}+\overrightarrow{Q_{4}^{0}}\cdot\overrightarrow{n_{4}}\chi _{i}\right\rangle
-\overrightarrow{i_{i}}\cdot\overrightarrow{E_{i}}
\end{equation}%
where $U_{i}$ denotes the volume internal energy of interfaces included in
the R.E.V..

Let us define $U_{\Sigma }$ as the sum of the volume internal energies of
the different phases

\begin{equation}
U_{\Sigma }=U_{3}+U_{4}+U_{i}
\end{equation}%
From (\ref{DerivMat}), we derive

\begin{equation}
\rho \frac{D}{Dt}\left( \frac{U_{\Sigma }}{\rho }\right)
=\sum\limits_{3,4}\left( \utilde{\sigma}_{k}:\utilde{grad}\overrightarrow{V_{k}}\right) +%
\overrightarrow{i}\cdot\overrightarrow{E}-div\overrightarrow{Q}  \label{Usigma}
\end{equation}%
where $\overrightarrow{i}$ represents the diffusion current, consisting of the diffusion
currents of the interfaces and of the cations in the solution

\begin{equation}
\overrightarrow{i}=\overrightarrow{I}-\sum\limits_{k=3,4}\left( \rho
_{k}Z_{k}\overrightarrow{V_{k}}\right) -Z_{i}\overrightarrow{V_{i}}=\rho
_{1}Z_{1}\left( \overrightarrow{V_{1}}-\overrightarrow{V_{4}}\right) +%
\overrightarrow{i_{i}}
\end{equation}

$U_{\Sigma }$ represents only a part of the internal energy of the whole
material; another part comes from the motion of the different phases in
the barycentric reference frame. The internal energy of the whole material
is defined by

\begin{equation}
U=E-E_{c}-E_{p}=U_{\Sigma }+E_{c\Sigma }-E_{c}
\end{equation}%
One deduces

\begin{equation}
\begin{tabular}{l}
$\rho \frac{D}{Dt}\left( \frac{U}{\rho }\right) =div\left(
\sum\limits_{3,4}E_{ck}\overrightarrow{V_{k}}-E_{c}\overrightarrow{V}\right)
+\frac{\partial }{\partial t}\left( \sum\limits_{3,4}E_{ck}-E_{c}\right) -div%
\overrightarrow{Q}$ \\ 
$\qquad \qquad \qquad \qquad \qquad \qquad \qquad \qquad
+\sum\limits_{3,4}\left( \utilde{\sigma}_{k}:%
\utilde{grad}\overrightarrow{V_{k}}\right) +%
\overrightarrow{i}\cdot\overrightarrow{E}$%
\end{tabular}%
\end{equation}

The first two terms in the right-hand side represent the volume internal
energy flux due to the relative velocities of the phases. The fourth one is
the volume kinetic energy converted into heat by viscous dissipation. And
the last term is the volume heat source by Joule effect in the solution.


\section{Discussion}

\label{sec:6} The conservation laws obtained for the global material include simplest cases. Assuming that the material is not electrically charged or removing the electric field, we obtain the equations governing a single-phase flow in porous medium \cite{Coussy95}. In case that the stress tensor is zero and that the velocities of the two phases are identical and uniform, we find the equations of a charged rigid solid subjected to an electric field.

The balance equations of the kinetic, potential, internal and
total\ energies all have the same structure : the energy variation following
the motion of one constituent, which is a particle derivative, is the sum of
a flux and of source terms. The equations we write are relative to a
thermodynamic closed system because of the use of the material derivative.
Source terms correspond to conversion of one kind of energy into another
one. At the microscopic scale, we obtain the following tables for the phase $%
k$

\begin{equation}
\begin{tabular}{cc}
\hline
& flux \\ \hline
$E_{pk}^{0}$ &  \\ 
$E_{ck}^{0}$ & $div\left( \utilde{\sigma}_{k}^{0}\cdot\overrightarrow{V_{k}^{0}}\right) $ \\ 
$U_{k}^{0}$ & $-div\overrightarrow{Q_{k}^{0}}$ \\ 
$E_{k}^{0}$ & $div\left( \utilde{\sigma}_{k}^{0}\cdot%
\overrightarrow{V_{k}^{0}}-\overrightarrow{Q_{k}^{0}}\right) $ \\ \hline
\end{tabular}%
\end{equation}%
and

\begin{equation}
\begin{tabular}{cccc}
\hline
& $E_{c}\longleftrightarrow E_{p}$ & $U\longleftrightarrow E_{p}$ & $%
E_{c}\longleftrightarrow U$ \\ \hline
$E_{pk}^{0}$ & $-\rho _{k}^{0}Z_{k}^{0}\overrightarrow{E_{k}^{0}}\cdot%
\overrightarrow{V_{k}^{0}}$ & $-\left( \overrightarrow{I_{k}^{0}}-\rho
_{k}^{0}Z_{k}^{0}\overrightarrow{V_{k}^{0}}\right)\cdot\overrightarrow{E_{k}^{0}%
}$ &  \\ 
$E_{ck}^{0}$ & $+\rho _{k}^{0}Z_{k}^{0}\overrightarrow{E_{k}^{0}}\cdot%
\overrightarrow{V_{k}^{0}}$ &  & $-\utilde{\sigma}
_{k}^{0}:\utilde{grad}\overrightarrow{V_{k}^{0}}$ \\ 
$U_{k}^{0}$ &  & $+\left( \overrightarrow{I_{k}^{0}}-\rho _{k}^{0}Z_{k}^{0}%
\overrightarrow{V_{k}^{0}}\right) \cdot\overrightarrow{E_{k}^{0}}$ & $+%
\utilde{\sigma}_{k}^{0}:\utilde{grad}\overrightarrow{V_{k}^{0}}$ \\ 
$E_{k}^{0}$ &  &  &  \\ \hline
\end{tabular}%
\end{equation}

Fluxes can be considered as the rate of variation of the quantity associated
with the conduction phenomenon. The flux of kinetic energy is due to the
contact force work, and the flux of internal energy to the heat conduction.
The total energy flux is then the sum of the two previous ones. We point out
that there is no flux for the potential energy. The viscous dissipation $%
\utilde{\sigma}_{k}^{0}:\utilde{grad}\overrightarrow{V_{k}^{0}}$ transforms the 
kinetic energy into heat, that is to say into internal energy. The work of the
electric forces produces two source terms : the first one is the
scalar product of the electric field $\overrightarrow{E_{k}^{0}}$ and of the
diffusion current $\overrightarrow{I_{k}^{0}}-\rho _{k}^{0}Z_{k}^{0}%
\overrightarrow{V_{k}^{0}}$, which is the electric current measured in the
barycentric reference frame. It can be seen as Joule heating, that is as a
conversion of potential energy into internal energy. The other part $\rho
_{k}^{0}Z_{k}^{0}\overrightarrow{V_{k}^{0}}\cdot\overrightarrow{E_{k}^{0}}$
results in a motion of the electric charges subject to the electric field;
potential energy is thus transformed into kinetic energy. Furthermore, the
energy conservation law is consequently satisfied. Accordingly, there is no
source term in the balance equation of the total energy.

We can examine in the same way the balance equations for one phase averaged
over the R.E.V. That highlights the source terms

\begin{equation}
\begin{tabular}{cccc}
\hline
& $E_{c}\longleftrightarrow E_{p}$ & $U\longleftrightarrow E_{p}$ & $%
E_{c}\longleftrightarrow U$ \\ \hline
$E_{pk}$ & $-\rho _{k}Z_{k}\overrightarrow{V_{k}}\cdot\overrightarrow{E_{k}}$ & $%
-\overrightarrow{i_{k}}\cdot\overrightarrow{E_{k}}$ &  \\ 
$E_{ck}$ & $+\rho _{k}Z_{k}\overrightarrow{V_{k}}\cdot\overrightarrow{E_{k}}$ & 
& $-\utilde{\sigma}_{k}:\utilde{grad}\overrightarrow{V_{k}}$ \\ 
$U_{k}$ &  & $+\overrightarrow{i_{k}}\cdot\overrightarrow{E_{k}}$ & $+%
\utilde{\sigma}_{k}:\utilde{grad}\overrightarrow{V_{k}}$ \\ \hline
\end{tabular}%
\end{equation}

Viscous dissipation and Joule heating transform kinetic energy and
potential energy into internal energy, respectively. And conversion of
potential energy into kinetic energy is due once more to electric charges
motion subject to the effect of the electric field. The other terms of the
equations can be presented in the form

\begin{equation}
\begin{tabular}{ccc}
\hline
& flux & interfacial exchanges \\ \hline
$E_{pk}$ &  &  \\ 
$E_{ck}$ & $div\left( \utilde{\sigma}_{k}\overrightarrow{V_{k}}\right) $ & $%
+\overrightarrow{F_{k}}\cdot\overrightarrow{V_{k}}$ \\ 
$U_{k}$ & $-div\overrightarrow{Q_{k}}$ & $-\left\langle \overrightarrow{%
Q_{k}^{0}}\cdot\overrightarrow{n_{k}}\chi _{i}\right\rangle $ \\ 
$E_{k}$ & $div\left( \utilde{\sigma}_{k}\cdot\overrightarrow{V_{k}}-%
\overrightarrow{Q_{k}}\right) $ & $+\overrightarrow{%
F_{k}}\cdot\overrightarrow{V_{k}}+P_{k}$ \\ \hline
\end{tabular}%
\end{equation}%
where

\begin{equation}
\begin{tabular}{l}
$\overrightarrow{F_{k}}\cdot\overrightarrow{V_{k}}=\left\langle \left( 
\utilde{\sigma}_{k}^{0}\cdot\overrightarrow{n_{k}}%
\right) \cdot\overrightarrow{V_{k}^{0}}\chi _{i}\right\rangle $ \\ 
$P_{k}=\left\langle -\overrightarrow{Q_{k}^{0}}\cdot\overrightarrow{n_{k}}\chi
_{i}\right\rangle $%
\end{tabular}%
\end{equation}

As before, the flux of internal energy is the heat transfer by conduction,
and the flux of kinetic energy is the volume power of the contact forces
within the phase. Additional\ terms arise from this analysis; they represent
exchanges between the phases through the interfaces. $\overrightarrow{F_{k}}\cdot%
\overrightarrow{V_{k}}$ is thus the volume power of the interaction forces
acting on the phase $k$ and corresponds to a kinetic energy input. $%
-\left\langle \overrightarrow{Q_{k}^{0}}\cdot\overrightarrow{n_{k}}\chi
_{i}\right\rangle $ results from the heat transfer through the interface and
modifies the internal energy. The sum of these two terms modifies the total
energy of the considered phase.

Concerning the whole E.A.P., we obtain the following decomposition

\begin{equation}
\begin{tabular}{cc}
\hline
& flux \\ \hline
$E_{p}$ &  \\ 
$E_{c}$ & $div\left[ \sum\limits_{3,4}\left( \utilde{%
\sigma}_{k}\cdot\overrightarrow{V_{k}}-E_{ck}\overrightarrow{V_{k}}\right)
+E_{c}\overrightarrow{V}\right] +\frac{\partial }{\partial t}\left(
E_{c}-\sum\limits_{3,4}E_{ck}\right) $ \\ 
$U$ & $div\left[ \sum\limits_{3,4}E_{ck}\overrightarrow{V_{k}}-E_{c}%
\overrightarrow{V}-\overrightarrow{Q}\right] +\frac{\partial }{\partial t}%
\left( \sum\limits_{3,4}E_{ck}-E_{c}\right) $ \\ 
$E$ & $div\left( \sum\limits_{3,4}\utilde{\sigma}_{k}%
\cdot\overrightarrow{V_{k}}-\overrightarrow{Q}\right) $ \\ \hline
\end{tabular}%
\end{equation}%
and

\begin{equation}
\begin{tabular}{cccc}
\hline
& $E_{c}\longleftrightarrow E_{p}$ & $U\longleftrightarrow E_{p}$ & $%
E_{c}\longleftrightarrow U$ \\ \hline
$E_{p}$ & $-\left( \sum\limits_{k=3,4}\left( \rho _{k}Z_{k}\overrightarrow{%
V_{k}}\right) +Z_{i}\overrightarrow{V_{i}}\right) \cdot\overrightarrow{E}$ & $-%
\overrightarrow{i}\cdot\overrightarrow{E}$ &  \\ 
$E_{c}$ & $+\left( \sum\limits_{3,4}\rho _{k}Z_{k}\overrightarrow{V_{k}}%
+Z_{i}\overrightarrow{V_{i}}\right) \cdot\overrightarrow{E}$ &  & $%
-\sum\limits_{3,4}\utilde{\sigma}_{k}:\utilde{grad}\overrightarrow{V_{k}}$ \\ 
$U$ &  & $+\overrightarrow{i}\cdot\overrightarrow{E}$ & $+\sum\limits_{3,4}%
\left( \utilde{\sigma}_{k}:\utilde{grad}\overrightarrow{V_{k}}\right) $ \\ \hline
\end{tabular}%
\end{equation}

The energy flux comes from the work of the contact forces in the different
phases and from the heat transfer by conduction; the first one is a flux of
kinetic energy, the second one is the flux of internal energy. The flux of
potential energy is still zero. An additional flux term appears : the
kinetic energy of the different phases measured in a barycentric reference
frame; this kinetic energy is indeed a part of the internal energy of the
global material. The source terms include viscous dissipation, which
transforms kinetic energy into heat, and Joule heating, which transforms
potential energy into internal energy. This last term is linked to the
diffusion current created by the interfacial charges motion and by the
cations motion in the solution reference frame. The global motion of the
charges under the influence of the electric field turns potential energy on
kinetic energy.


\section{Conclusion}

\label{sec:concl} We have modelled an electroactive, ionic, water-saturated
polymer placed in an electric field. The polymer is fully dissociated,
releasing cations of small size. This system is depicted as the
superposition of two continuous media : a deformable porous medium
constituted by the polymer backbone embedded with anions, in which flows an
ionic solution composed by water and released cations. We have deduced the
microscale conservation laws of each phase : mass continuity equation,
linear momentum conservation law, Maxwell's equations and energy balance
laws. Then we derived the physical quantities attached to the interfaces. An
average over the R.E.V. of the material has provided one with macroscale
conservation laws for each phase first and for the global E.A.P., next.
Having the three constituents of the material (solid, solvent and cations)
different velocities, we have used for this last step, the material
derivative concept in order to obtain an Eulerian formulation of the
conservation laws.

We have examined the balance equations of the different energies (kinetic,
potential and internal ones), and we have put the emphasis on the phenomena
responsible for the conversion of one kind of energy into another : viscous
frictions, Joule effect and charge motion under the effect of the electric
field. The first two results in dissipation. Moreover, the macroscale
equations relative to each phase allow an evaluation of energy exchanges
through the interfaces.

Using the linear thermodynamics of the irreversible processes we should now
be able to determine the potential of dissipation and to derive the
phenomenological equations governing this system. This will be the subject of a forthcoming work.


\section{Appendix : interface modelling}

\label{sec:Annexe} In practice, contact area between phases $3$ and $4$ has
a certain thickness; extensive physical quantities like mass density, linear
momentum and energy continuously vary from one bulk phase to the other one.
This complicated reality can be modelled by two uniform bulk phases
separated by a discontinuity surface $\Sigma $ whose localization is
arbitrary. Let $\Omega $ be a cylinder crossing $\Sigma $, whose bases are
parallel to $\Sigma $. We denote by $\Omega _{3}$ and $\Omega _{4}$ the parts
of $\Sigma $ respectively included in phases $3$ and $4$.

The continuous quantities relative to the contact zone are identified by a
superscript $^{0}$ and no subscript. A microscale quantity per surface unit $%
g_{i}^{0}$ related to the interface is defined by

\begin{equation}
g_{i}^{0}=\lim\limits_{\Sigma \longrightarrow 0}\frac{1}{\Sigma }\left\{
\int_{\Omega }g^{0}dv-\int_{\Omega _{3}}g_{3}^{0}dv-\int_{\Omega
_{4}}g_{4}^{0}dv\right\}   \label{Def-i0}
\end{equation}%
where $\Omega _{3}$ and $\Omega _{4}$ are small enough so that $g_{3}^{0}$
and $g_{4}^{0}$ are constant. Its average over the R.E.V. is the volume
quantity $g_{i}$ defined by

\begin{equation}
g_{i}=\left\langle g_{i}^{0}\chi _{i}\right\rangle 
\end{equation}%
The balance equation of the interfacial quantity $g_{i}^{0}$ is written as (Ishii, 
\cite{Ishii06})%
\[
\frac{\partial g_{i}^{0}}{\partial t}+div_{s}\left( g_{i}^{0}\overrightarrow{%
V_{i}^{0}}\right) =\sum\limits_{3,4}\left[ g_{k}^{0}\left( \overrightarrow{%
V_{k}}-\overrightarrow{V_{i}^{0}}\right) \cdot\overrightarrow{n_{k}}+%
\overrightarrow{J_{k}^{0}}\cdot\overrightarrow{n_{k}}\right] -div_{s}%
\overrightarrow{J_{i}^{0}}+\phi _{i}^{0}
\]%
where $div_{s}$ denotes the surface divergence operator. $\overrightarrow{%
J_{i}^{0}}$ is the surface flux of $g_{i}^{0}$, $\overrightarrow{%
J_{k}^{0}}$ the flux of $g_{k}^{0}$ and $\phi _{i}^{0}$ the
surface source term. 

We arbitrarily fix the interface position in such a way that it has no mass
density

\begin{equation}
\rho _{i}^{0}=\lim\limits_{\Sigma \longrightarrow 0}\frac{1}{\Sigma }\left\{
\int_{\Omega }\rho ^{0}dv-\int_{\Omega _{3}}\rho _{3}^{0}dv-\int_{\Omega
_{4}}\rho _{4}^{0}dv\right\} =0  \label{DefI}
\end{equation}%
From (\ref{CMcl}), we deduce that the linear momentum and the kinetic energy
per surface unit of the interface, respectively denoted $\overrightarrow{%
P_{i}^{0}}$ and $E_{ci}^{0}$, are zero

\begin{equation}
\overrightarrow{P_{i}^{0}}=\overrightarrow{0}\qquad \qquad E_{ci}^{0}=0
\end{equation}

In the same way, we define the charge per unit surface $Z_{i}^{0}$, the
surface current vector $\overrightarrow{I_{i}^{0}}$, the surface diffusion
current $\overrightarrow{i_{i}^{0}}$, the surface potential energy $E_{pi}^{0}$, 
the surface internal energy $U_{i}^{0}$ and the surface total energy $E_{i}^{0}$.

The balance equations of these quantities write%
\begin{equation}
\frac{\partial Z_{i}^{0}}{\partial t}+div_{s}\left( Z_{i}^{0}\overrightarrow{%
V_{i}^{0}}\right) =\overrightarrow{i_{3}^{0}}\cdot\overrightarrow{n_{3}}+%
\overrightarrow{i_{4}^{0}}\cdot\overrightarrow{n_{4}}-div_{s}\overrightarrow{%
i_{i}^{0}}
\end{equation}

\begin{equation}
\frac{\partial \overrightarrow{P_{i}^{0}}}{\partial t}+\overrightarrow{div_{s}}%
\left(\overrightarrow{P_{i}^{0}}\otimes \overrightarrow{V_{i}^{0}}\right) =-%
\utilde{\sigma}_{3}^{0}\cdot\overrightarrow{n_{3}}-%
\utilde{\sigma}_{4}^{0}\cdot\overrightarrow{n_{4}}%
+Z_{i}^{0}\overrightarrow{E_{i}^{0}}
\end{equation}

\begin{equation}
\frac{\partial E_{pi}^{0}}{\partial t}+div_{s}\left( E_{pi}^{0}%
\overrightarrow{V_{i}^{0}}\right) =-\overrightarrow{I_{i}^{0}}\cdot%
\overrightarrow{E_{i}^{0}}
\end{equation}

\begin{equation}
\frac{\partial E_{i}^{0}}{\partial t}+div_{s}\left( E_{i}^{0}\overrightarrow{%
V_{i}^{0}}\right) =-\left( \utilde{\sigma}_{3}^{0}\cdot%
\overrightarrow{n_{3}}\right) \cdot\overrightarrow{V_{3}^{0}}-\left( 
\utilde{\sigma}_{4}^{0}\cdot\overrightarrow{n_{4}}%
\right) \cdot\overrightarrow{V_{4}^{0}}+\overrightarrow{Q_{3}^{0}}\cdot%
\overrightarrow{n_{3}}+\overrightarrow{Q_{4}^{0}}\cdot\overrightarrow{n_{4}}
\end{equation}

\begin{equation}
\frac{\partial U_{i}^{0}}{\partial t}+div_{s}\left( U_{i}^{0}\overrightarrow{%
V_{i}^{0}}\right) =\overrightarrow{Q_{3}^{0}}\cdot\overrightarrow{n_{3}}+%
\overrightarrow{Q_{4}^{0}}\cdot\overrightarrow{n_{4}}+%
\overrightarrow{i_{i}^{0}}\cdot\overrightarrow{E_{i}^{0}}
\end{equation}
Averaging over the R.E.V., this leads to the boundary conditions below

\begin{equation}
\frac{\partial Z_{i}}{\partial t}+div\overrightarrow{I_{i}}=\left\langle 
\overrightarrow{i_{3}^{0}}\cdot\overrightarrow{n_{3}}\chi _{i}\right\rangle
+\left\langle \overrightarrow{i_{4}^{0}}\cdot\overrightarrow{n_{4}}\chi
_{i}\right\rangle 
\end{equation}

\begin{equation}
\overrightarrow{F_{3}}+\overrightarrow{F_{4}}=Z_{i}\overrightarrow{E_{i}}
\end{equation}%
\begin{equation}
\frac{\partial E_{pi}}{\partial t}+div\left( E_{pi}\overrightarrow{V_{i}}%
\right) =-\overrightarrow{I_{i}}\cdot\overrightarrow{E_{i}}
\end{equation}

\begin{equation}
\frac{\partial E_{i}}{\partial t}+div_{s}\left( E_{i}\overrightarrow{V_{i}}%
\right) =-P_{3}-P_{4}-\overrightarrow{F_{3}}\cdot\overrightarrow{V_{3}}-%
\overrightarrow{F_{4}}\cdot\overrightarrow{V_{4}}
\end{equation}%
\begin{equation}
\frac{\partial U_{i}}{\partial t}+div\left( U_{i}\overrightarrow{V_{i}}%
\right) =\left\langle \overrightarrow{Q_{3}^{0}}\cdot\overrightarrow{n_{3}}\chi
_{i}+\overrightarrow{Q_{4}^{0}}\cdot\overrightarrow{n_{4}}\chi _{i}\right\rangle
+\overrightarrow{i_{i}}\cdot\overrightarrow{E_{i}}
\end{equation}

Moreover, we have%
\[
\overrightarrow{I_{i}}=Z_{i}\overrightarrow{V_{i}}+\overrightarrow{i_{i}}
\]

\section{Notations}

\label{sec:not} $k=1,2,3,4,i$ subscripts respectively represent cations,
solvent, solid, solution (water and cations) and interface; quantities
without subscript refer to the whole material. Superscript $^{0}$ denotes a
local quantity; the lack of superscript indicates average quantity at the
macroscopic\ scale. Microscale volume quantities are relative to the volume
of the phase, average quantities to the volume of the whole material.

\begin{description}
\item $C$ : cations molar concentration (relative to the liquid phase);

\item $\overrightarrow{D}$ ($\overrightarrow{D_{k}}$,$\overrightarrow{%
D_{k}^{0}}$) : electric displacement field;

\item $E$ ($E_{k}$,$E_{k}^{0}$) : total energy density (internal, kinetic
and potential);

\item $\overrightarrow{E}$ ($\overrightarrow{E_{k}}$,$\overrightarrow{%
E_{k}^{0}}$) : electric field;

\item $E_{c}$ ($E_{c\Sigma }$,$E_{ck}$,$E_{ck}^{0}$) : kinetic energy
density;

\item $E_{p}$ ($E_{pk}$,$E_{pk}^{0}$) : potential energy density;

\item $F=96487\;C\;mol^{-1}$ : Faraday's constant ;

\item $\overrightarrow{F_{k}}$ : resultant of the mechanical stresses
exerted on the phase $k$ by the other phase;

\item $\overrightarrow{I}$ ($\overrightarrow{I_{k}}$,$\overrightarrow{%
I_{k}^{0}}$) : current density vector;

\item $\overrightarrow{i}\ $($\overrightarrow{i_{k}}$,$\overrightarrow{%
i_{k}^{0}}$) : diffusion current;

\item $M_{k}$ : molar mass of component $k$;

\item $\overrightarrow{n_{k}}$ : outward-pointing unit normal of phase $k$;

\item $P_{k}$ : heat flux through interfaces;

\item $\overrightarrow{P_{i}^{0}}$ : local surface linear momentum of interface;

\item $\overrightarrow{Q}$ ($\overrightarrow{Q_{k}}$,$\overrightarrow{%
Q_{k}^{0}}$) : heat flux;

\item $U$ ($U_{\Sigma }$,$U_{k}$,$U_{k}^{0}$) : internal energy density;

\item $\overrightarrow{V}$ ($\overrightarrow{V_{k}}$,$\overrightarrow{%
V_{k}^{0}}$) : velocity;

\item $z_{k}$ : number of elementary charges of a ion $k$;

\item $Z$ ($Z_{k}$,$Z_{k}^{0}$) : total electric charge per unit of mass;

\item $Z_{i}$ ($Z_{i}^{0}$) : electric charge density per unit surface;

\item $\varepsilon $ ($\varepsilon _{k}$,$\varepsilon _{k}^{0}$) :
permittivity;

\item $\rho $ ($\rho _{k}$,$\rho _{k}^{0}$) : mass density;

\item $\utilde{\sigma }$ ($\utilde{\sigma}_{k}$,$\utilde{\sigma}_{k}^{0}$) : stress tensor;

\item $\phi _{k}$ : volume fraction of phase $k$;

\item $\chi _{k}$ : function of presence of phase $k$ ;
\end{description}

\begin{acknowledgements}
The authors would like to thank D. Lhuillier and O. Kavian for their fruitful and stimulating discussions.
\end{acknowledgements}



\end{document}